# Design and Comparison
# of a 1 MW / 5s HTS SMES
# with Toroidal and Solenoidal Geometry

Antonio Morandi, *Senior Member, IEEE*, Massimo Fabbri, Babak Gholizad, Francesco Grilli, Frédéric Sirois, *Senior Member, IEEE* and Víctor M. R. Zermeño

*Abstract*— The design of a HTS SMES coil with solenoidal and toroidal geometry is carried out based on a commercially available 2G HTS conductor. A SMES system of practical interest (1 MW / 5 s) is considered. The comparison between ideal toroidal and solenoidal geometry is first discussed and the criteria used for choosing the geometrical parameters of the coils' bore are explained. The design of the real coil is then carried out and the final amount of conductor needed is compared. A preliminary comparison of the two coils in terms of AC loss during one charge discharge cycle is also discussed.

## I. INTRODUCTION

TOROIDAL AND SOLENOIDAL geometries can be considered for the design of SMES coils. The solenoid is simpler to manufacture and allows an easier handling of the mechanical stress. Moreover, for isotropic superconductors, the solenoid allows minimum wire consumption and represents the most cost effective solution [1]-[2]. Despite the drawback of the high stray field, solenoidal geometry has been exploited in the past for the development of real-scale SMES systems based on low temperature superconductors [3]-[4]. More recently solenoidal geometries have also been used for the development of SMES based on first generation HTS wires [5]-[6]. However, when 2G HTS materials are considered, toroidal geometry is generally [7]-[10]. Since toroid minimizes the perpendicular component of the magnetic field on the conductor, a lower material requirement can be expected due to the drastic dependence of its $J_c$ vs $B$ performance on the orientation of the field. A presumably lower level AC loss can also be expected due to the lower perpendicular field. It is to be considered, however, that a careful investigation is needed in order to assess the conductor requirement and the AC loss of real toroidal coils with reduced, but non-negligible, perpendicular field.

In this paper the design of a HTS SMES coil with solenoidal and toroidal geometry is carried out. In both cases the coil is made of multiple identical pancakes [5]-[11]. A SMES system of practical interest (1 MW / 5 s) is considered. The main characteristics are summarized in section II. The design of the coil is carried out based on a commercially available 2G HTS conductor which is introduced in section III. The comparison between ideal toroidal and solenoidal geometry is first discussed in section IV and the criteria used for choosing the geometrical parameters of the coils' bore are explained. The design of the real coil is then carried out in section V. The general characteristics are presented and the final amount of conductor needed is calculated and compared. Finally, in section VI, a preliminary comparison of the two windings in terms of AC loss during one charge discharge cycle is presented.

## II. MAIN CHARACTERISTICS OF THE SMES SYSTEM

A SMES able to deliver 1 MW for 5 s is considered. The main characteristics of the system are listed in Table 1. This SMES can be used for several applications in industry including the protection of critical loads from voltage sag, leveling of large impulsive load and for bridging the onset time of off-line power reserves like, for example, diesel generators [12]. A voltage of 1 kV is chosen for the DC bus, which is below the insulation level achievable for conduction-cooled coils and is typical for the MW level power electronics. As a consequence, since the aim is to provide 1 MW power, the minimum current $I_{min}$ of the SMES cannot be below 1 kA during the delivery interval. If the SMES is charged or discharged with constant power $P$ during a certain time interval $\Delta t$ the current $I$ at the end of this interval is given by

$$I = \sqrt{I_0^2 \pm 2P\Delta t/L} \qquad (1)$$

where $I_0$ is the current at the begin of the interval and $L$ is the inductance of the coil. The '+' and '−' signs apply for the charge and the discharge respectively. Equation (1) points out that the lower the inductance chosen for the SMES the higher the current that is reached at the end of the charging phase. A maximum current $I_{max}$ of 3 kA is assumed here in order to keep heat load and size of current leads within acceptable values and to reduce the current rating of the DC/DC switches. An inductance of 1.25 H is obtained for the coil based on the

Antonio Morandi, Massimo Fabbri and Babak Gholizad are with the University of Bologna - Dept. of Electrical, Electronic and Information Engineering, Viale Risorgimento n. 2, 40136 Bologna, Italy

Francesco Grilli and Victor Manuel Rodrigo Zermeño are with Karlsruhe Institute of Technology, Institute for Technical Physics (ITEP), Hermann-von-Helmholtz-Platz 1, D-76344 Eggenstein-Leopoldshafen

Frédéric Sirois is with Département de Génie Électrique, École Polytechnique de Montréal, Montréal, QC, Canada

Corresponding author: Antonio Morandi (antonio.morandi@unibo.it)

deliverable power and interval (see Table I) and the values chosen for $I_{min}$ and $I_{max}$. Finally, a maximum stored energy $E$ of 5.6 MJ is obtained for the coil corresponding to $I_{max}$.

TABLE I
MAIN DATA OF THE SMES SYSTEM

| | |
|---|---|
| Deliverable Power, $P$ | 1 MW |
| Duration of delivery, $\Delta t$ | 5 s |
| Deliverable Energy, $\Delta E$ | 5 MJ |
| Voltage of the DC bus, $V_{dc}$ | 1 kV |
| Minimum Current, $I_{min}$ | 1 kA |
| Maximum Current, $I_{max}$ | 3 kA |
| Inductance, $L$ | 1.25 H |
| Stored Energy at $I_{max}$, $E$ | 5.6 MJ |

## III. HTS CONDUCTOR AND CABLE

A commercial 2G HTS conductor produced by Superpower is considered for the design of the SMES. This consists of a tape with 2×20 μm copper stabilization and 0.1 mm total thickness [13]. The engineering critical current density of the tape at 77 K and self-field operation $J_c(77 K, sf)$ is $3 \cdot 10^8$ A/m$^2$. The dependence of the critical current density on the parallel ($B_\parallel$) and perpendicular ($B_\perp$) components of the applied field at 20 K can be expressed as

$$J_C(20K, B_\parallel, B_\perp) = L_0 \frac{J_C(77K, \text{sf})}{(1 + \sqrt{k^2 B_\parallel^2 + B_\perp^2}/B_0)^b} \quad (2)$$

with $L_0 = 6.52$; $k = 0.06$; $B_0 = 5.04\,T$; $b = 1.64$;

Parameters $L_0$, $k$, $B_0$ and $b$ of equation (1) have been obtained from the fitting of experimental data. More information on the $J_c$ versus $B$ performance of the tape and its fitting law in a wider range of temperature can be found in [14]. Since the maximum required current capacity is 3 kA several conductors must be assembled in parallel for the manufacturing of the SMES coil. A Roebel cable [15] is considered for this purpose. This is made of a number HTS strands continuously transposed. The cable is wrapped with 125 μm electrical insulation. The width of each strand is 5.5 mm. The total wide side of the cable is 12 mm (a separation of 1 mm is left between strands along the wide side of the cable for manufacturability reasons). A filling factor of 0.76, which is typical for practical Roebel cables, is considered in section V for the design of the coils.

## IV. SIZING OF TOROIDAL AND SOLENOIDAL BORE

In order to carry out the detailed design of the SMES coil, the geometrical parameters of the coil's bore must first be fixed. This means that some criteria need to be introduced for choosing toroidal and poloidal radius of the torus and height and inner radius of the solenoid. A reference level of magnetic flux density, which is indicated with $B_{ref}$ in the following, needs also to be specified. Concerning the toroid we assume as reference value $B_{ref}$ the maximum flux density at the innermost part of the torus. Given the toroidal radius $R_T$ and the poloidal radius $R_P$ the aspect ratio $a$ of the torus can be defined that is the ration between the toroidal circumference and the poloidal diameter ($a = \pi R_T/R_P$). The energy $E$ of an ideal torus toroid with toroidal radius $R_T$, reference flux density $B_{ref}$ and aspect ratio $a$ is expressed as

$$E = \frac{2\pi^2}{\mu_0} B_{ref}^2 R_T^3 (1 - \pi/a)^2 \left(1 - \sqrt{1 - \pi^2/a^2}\right) \quad (1)$$

Equation (1) is obtained by integrating the magnetic energy density $e = B^2/2\mu_0$ over the volume of the torus and by considering the dependence of the flux density $B$ on the radial coordinate. More details for the deduction of (1) are given in the appendix. If the stored energy $E$ is to be achieved with a torus of given aspect ratio $a$ and reference flux density $B_{ref}$, then a unique toroidal radius $R_T$ exists that can be obtained from (1). The overall diameter $D_T$ of this torus can be defined as $D_T = 2R_T(1 + \pi/a)$. In Fig. 1 the total size of the toroid with stored energy 5.6 MJ is plotted as function of the aspect ratio $a$ for different values of the reference field. As it can be seen that, independently of the field level, a unique value of the aspect ratio $a = 13\pi/5 \approx 8.2$ exists which minimizes the overall size of the toroid (see the appendix for more details on the deduction of this value). We have verified that this holds also in case of different values of the stored energy. In section V an aspect ratio $a = 8.2$ will be considered for the design of a real toroidal coil made of multiple pancakes.

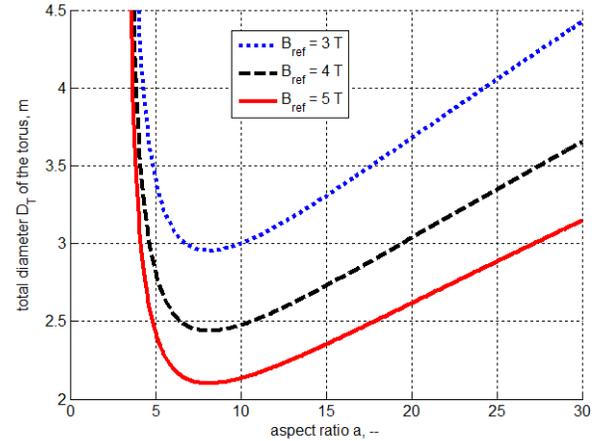

Fig.1. Total size of the ideal toroid with stored energy 5.6 MJ versus the aspect ratio $a$ for different values of the reference field.

As for the solenoid, we assume as a first approximation that the magnetic field does not change substantially within the cylindrical coil's bore. With this assumption the energy $E$ stored in a solenoid with inner radius $R_S$, height $H_S$ and magnetic flux density $B_{ref}$ can be approximated by $E \approx \pi R_S^2 H_S B_{ref}^2 / 2\mu_0$. In the following we assume for the solenoid a height equal to the overall size of the toroid, which is $H_S = D_T$. The radius is obtained accordingly from the approximated expression of the energy.

## V. Design of the SMES coils

The design of a SMES coil made of multiple pancakes is dealt with in this section. The pancakes can be stacked in the axial direction in order to form a solenoid [5], [6], [11] or they can be arranged along the toroidal circumference in order to form a torus [7], [10]. Each of the pancakes is made of multiple turns of Roebel cable. Two copper plates with 1 mm thickness are placed on each side of the pancakes in order to allow thermal connection with the cooling system. A $J/J_c$ factor of 0.6 is considered for the design of the coil. A reference field level $B_{ref}$ = 4 T is assumed. This is a compromise value. Higher values of $B_{ref}$ require more conductor [12]. Lower values reduce tape consumption but increase the overall dimension of the coil. However, the results presented in the following in terms of comparison between toroidal and solenoidal coils also apply for different field value.

The procedure described in section IV is first used for choosing toroidal and poloidal radius of the toroidal coil and height and inner radius of the solenoidal one. The number of pancakes of each of the coils is deduced from the ratio between the available space (height of the solenoid or innermost circumference of the toroid) and the gross height of the pancakes (including the insulated Roebel cable and the copper plates). In the case of torus, an additional space between the pancakes (needed for manufacturability) is also taken into account. For a given thickness of the pancakes, the maximum operating current density $J_{max}$ of the coil is found by intersecting the coil's load line (scaled by the inverse of the filling factor) with the $J_c$ vs $B$ curve of the material (scaled by the $J/J_c$ factor). The stored energy $E_{max}$ corresponding to $J_{max}$ is also calculated. The final layout is obtained by choosing for the pancakes the minimum thickness that allows reaching the required value of stored energy ($E_{max}$ = 5.6 MJ). Finally, each of the pancakes is then subdivided in as many turns as needed for obtaining an overall inductance $L$ = 1.25 H for the coil in order to meet the requirement of deliverable energy ($\Delta E$ = 5 MJ), as specified in section II. This subdivision only changes the inductance but does not affect the field distribution and the stored energy.

The main characteristics of the toroidal and solenoidal SMES coils are listed in Table II. The field map at the maximum operating current is shown in Fig. 2. Only one quarter of the coils is shown. Lines of constant field outside the coil's bore are also plotted for the solenoid. The arrangement of the pancakes is also visible in the figure. It can be deduced that, despite the lower perpendicular component of the field that occurs on the HTS tape (see next section for a quantitative discussion on this aspect), a relatively high amount of conductor is needed for the toroid (59.8 km, 12 % lower than 68.1 km required for the solenoid). This is due to the decrease of the field with the radial coordinate, which requires a larger field volume, and hence a larger poloidal radius, compared to the solenoid for achieving the same storage capacity. A larger overall size is obtained with the toroid, which implies a much larger heat load due to radiation. In fact, the area of the cylindrical envelop of the torus is about 300% that of the solenoid. Furthermore, a more complex support structure is also needed for the pancakes, which increases the manufacturing costs.

The drawback of the solenoid is the much larger stray field, which can have negative effects on the cryocoolers and power electronics. In fact, in case of toroid, the line with constant stray field at 50 mT, which is assumed as practical limit in industry for the shielding of sensitive apparatus, is in the immediate vicinity of the outermost part of the coil (not visible in Fig. 1) and no shielding is required. In case of solenoid, the 50 mT line extends far from the bore and shielding is needed. However, since moderate field level (< 500 mT) is obtained, passive shielding can be implemented by means of magnetic steel placed in the vicinity of the vacuum chamber. Laminated materials can be used in order to reduce additional losses due to induced currents.

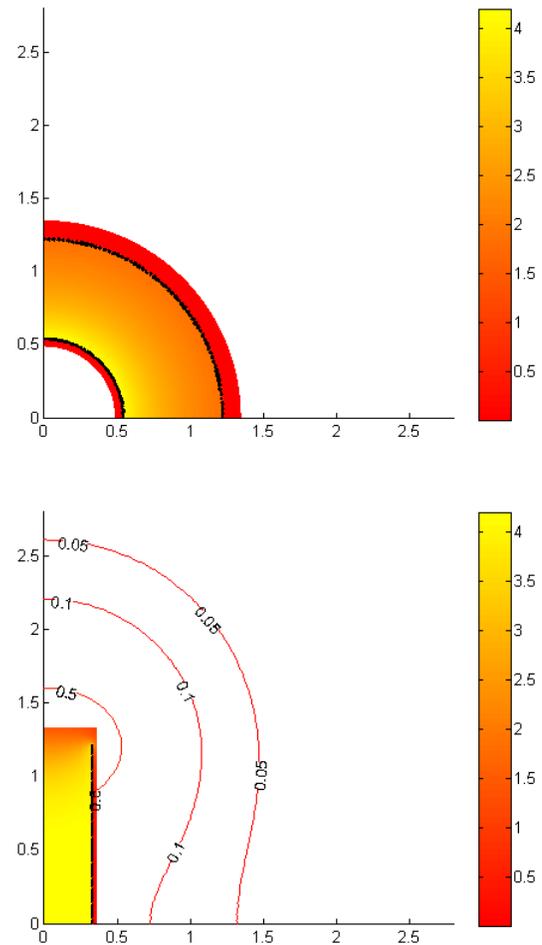

Fig.2. Field map of the toroidal (top) and the solenoidal (bottom) SMES. Only one quarter of the coils is shown. Lines of constant stray field (at 500 mT, 100 mT and 50 mT) are also plotted for the solenoid.

TABLE II
MAIN CHARACTERISTICS OF THE REAL TOROIDAL AND SOLENOIDAL SMES
SYSTEM FULFILLING THE REQUIREMENTS OF TABLE I

|  | Toroid | Solenoid |
|---|---|---|
| Poloidal radius / Inner radius | 338 mm | 320 mm |
| Toroidal radius / Height | 882 mm | 2425 mm |
| Number of pancakes | 236 | 183 |
| Thickness of one pancake | 6.9 mm | 10.6 mm |
| Number of turns per pancake | 16 | 16 |
| Number of HTS conductors per cable | 10 | 16 |
| Max perp. field on the conductor at $I_{max}$ | 1.61 T | 3.76 T |
| Max parallel field on the Conductor at $I_{max}$ | 4.36 T | 4.31 T |
| Total length of HTS tape of the coil | 59.8 km | 68.1 km |

## VI. AC LOSS

Advanced numerical techniques need to be employed for calculating the total loss of the SMES coils during a complete charge/discharge cycle. This is particularly true for the toroid geometry, where the 2D assumption may be inaccurate due to the fact that the perpendicular component of the magnetic field changes along the longitudinal direction of the tape. Further complications, which apply also in case of solenoid, come from the large size of the coil and the intrinsic 3D behavior of the Roebel cable. Furthermore, possible losses due to eddy currents in the normal conducting matrix and additional losses in metallic parts (e.g. copper plates or braids for the thermal connection of the coil with the cryocoolers) must be taken into account. No attempt is made in the following to provide loss data for the considered SMES coils. However, some quantitative investigation is carried out, under simplifying assumptions, with the aim to compare the toroidal and the solenoidal coil in terms of loss.

In Fig. 3 the length of conductor is plotted versus the perpendicular field to which it is subjected when the SMES operates at the maximum current. Data shown were obtained by first subdividing each of the tapes of the coil in small portions, then by calculating the perpendicular field acting on each of them and finally by summing up for obtaining the length of conductor subject to a field falling within a given interval. The field range from 0 to 4 T was analyzed by subdivision in intervals of 0.1 T each. By processing the data of Fig. 3 we note that in case of solenoid 50.8 km of tape (74.7% of the total) experience a perpendicular field below or equal to 0.5 T, whereas in case of toroid the amount of conductor within the same field interval is 11.2 km. Furthermore, the amount of tape subject to a perpendicular field in the range 1.0 – 1.6 T is 5.1 km for the solenoid and 37.1 km for the toroid. Based on this data we see that, differently from the ideal torus where no perpendicular field exists and hence negligible loss occur, substantial losses can be expected for the real toroidal coil due to the substantial perpendicular field, which involves a large amount of conductor. The maximum perpendicular field for the torus is 1.61 T. Much higher levels of perpendicular field, up to 3.76 T, exist for the solenoid but these involve 3.4 km (5%) of conductor only. In particular, only 0.2 km of conductor are exposed to a very high field in the 3.0 – 4.0 T range.

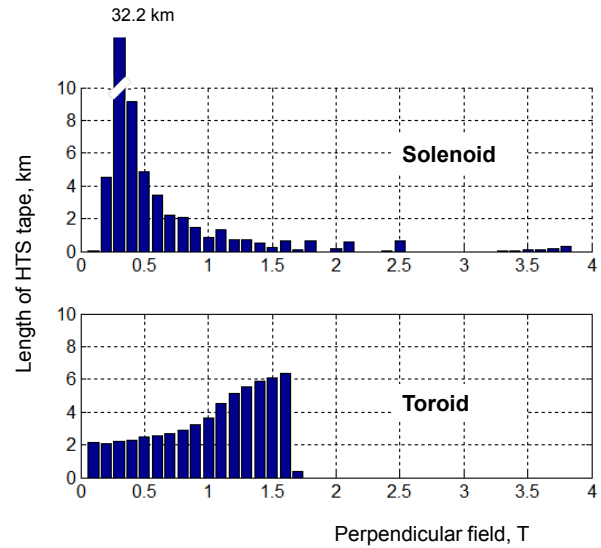

Fig.3. Length of conductor versus the perpendicular field to which it is subjected when the SMES operates at the maximum current. The field range is 0 – 4 T and is subdivided in intervals of 0.1 T. For example, the highest bar of the top graph shows that 32.2 km of tape are subject to a field in the internal 0.2 – 0.3 T in the solenoid.

In order to compare the performance of the toroidal and the solenoidal coil in terms of AC losses we have calculated the loss per unit length of one single HTS conductor with 5.5 mm width subject to the field produced by the whole coil during one discharge/charge cycle. The perpendicular field only is considered. The transport current is also applied to the tape. During the discharge/charge cycle the transport current of the Roebel cable change from 3 kA to 1 kA and then back to 3 kA according to equation (1). The transport current applied to the conductor is obtained from the transport current of the cable divided by the number of conductors of the cable (see table II). A peak transport current of 187.5 A and 300 A is obtained for the solenoid and the toroid respectively. The AC losses were computed with a finite-element model based on the 2-D H-formulation of Maxwell's equation [16]. Results are shown in Fig 4. For the same field level, only slightly higher losses per cycle per unit length of conductor are obtained for the toroid, despite the much higher transport current. This confirms that the losses are dominated by the applied field. As a figure of merit for comparing the loss performance of the two coils we have combined the loss per unit length corresponding to each of the levels of perpendicular field with the length of conductor exposed to that field, thus obtaining a quantitative indicator which we refer to as "loss indicator". In particular, we have multiplied the loss per cycle obtained for various field (see Fig. 3 ) times the corresponding length of conductor (see Fig. 4) and we have summed up all the contributions. Results are reported in table III. We see that greater loss can be expected for the toroid. We stress again, however, that this data do not represent an estimation of the total loss of the SMES coil. We also point out that a difference of the solenoid is that filamentarization, which is very likely to be needed for reducing the AC loss of SMES coil in regions of high perpendicular field, only needs to be applied for few pancakes at the ends of the coil. In contrast, in case of toroid, filamentarization must be applied to all the length of the conductor since the perpendicular field is distributed evenly all over the pancakes (see Fig. 3).

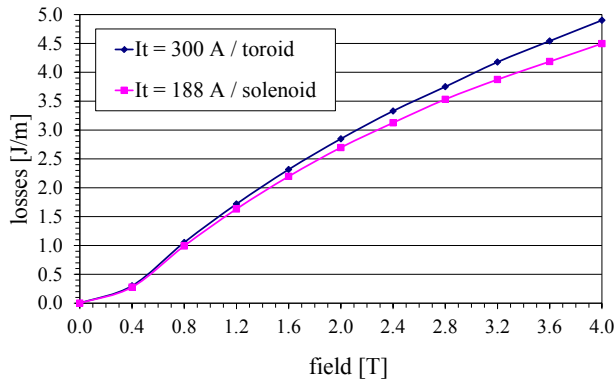

Fig.4. AC loss of one HTS tape with 5.5 mm width subject to perpendicular field and transport current.

TABLE III
LOSS INDICATOR OF THE SMES COILS

|  | Solenoid |
|---|---|
| Loss indicator of the toroid | 76667 J |
| Loss indicator of the solenoid | 31032 J |

## VII. CONCLUSION

A unique value of the aspect ratio exists that minimizes the overall size of a toroidal coil. Differently than for the ideal case, a substantial perpendicular field occurs on the HTS conductor for a real toroidal coil. A much higher perpendicular field occurs, however, in a solenoidal coil with same stored energy and maximum size, which lead to lower critical current of the HTS conductor. Nevertheless, based on actual $J_c$ vs $B$ performance of present state-of-the-art HTS materials, comparable lengths of conductor are needed for the solenoid and the toroidal coils. This is due to the fact that for the toroid, the field decreases with the radial coordinate and larger volume is needed to store the same energy. Moreover, due to the fact that the (lower) perpendicular field is more evenly distributed on the conductor, greater AC loss can be expected for the toroid compared to the solenoid.

## APPENDIX. ENERGY AND TOTAL SIZE OF AN IDEAL TOROIDAL COIL

An ideal torus with toroidal radius $R_T$ and aspect ratio $a$ is considered. If $B_{ref}$ is the flux density at the innermost part of the torus the distribution within the torus volume is given by

$$B(R) = B_{ref} \frac{(R_T - R_P)}{R} \quad (3)$$

The magnetic energy E can be obtained by integrating the density $e = B^2/2\mu_0$ over the volume of the torus that is .

$$E = \frac{2\pi B_{ref}^2}{\mu_0} \int_{R_T(1-\pi/a)}^{R_T(1+\pi/a)} \frac{R_T^2(1-\pi/a)^2}{R} \sqrt{R_T^2(1-\pi/a) - (R-R_T)^2}\, dR \quad (4)$$

By using the substitution $R = S\, R_T$ we obtain:

$$E = \frac{2\pi}{\mu_0} B_{ref}^2 R_T^3 (1-\frac{\pi}{a})^2 \int_{(1-\frac{\pi}{a})}^{(1+\frac{\pi}{a})} \frac{1}{S} \sqrt{\left(\frac{\pi}{a}\right)^2 - (S-1)^2}\, dS \quad (5)$$

By solving analytically the integral in (5) [17], we finally obtain equation (2). If the stored energy $E$ is to be achieved with a torus of given aspect ratio $a$ and reference flux density $B_{ref}$, then a unique toroidal radius $R_T$ exists that can be obtained from (1). The overall diameter $D_T$ of this torus is $D_T = 2R_T(1 + \pi/a)$ and can be expressed as

$$D_T = \left(\frac{4\mu_0 E}{\pi^2 B_{ref}^2}\right)^{\frac{1}{3}} (1+\pi/a) \left[(1-\pi/a)^2 \left(1-\sqrt{1-\pi^2/a^2}\right)\right]^{-\frac{1}{3}} \quad (5)$$

By analyzing the derivative it can be found, after some manipulation, that function $D_T(a)$ of equation (6) has a minimum at $a = 13\pi/5$. Note that this minimum is independent on $E$ and $B_{ref}$.